\documentstyle[11pt]{article}
\begin{document}
\renewcommand{\theequation}{\thesection.\arabic{equation}}
\renewcommand{\refname}{References.}
\newcommand{\sect}[1]{ \section{#1} \setcounter{equation}{0} }
\newcommand{\partialslash}{\partial \! \! \! /}
\newcommand{\xslash}{x \! \! \! /}
\newcommand{\yslash}{y \! \! \! /}
\newcommand{\zslash}{z \! \! \! /}
\newcommand{\half}{\mbox{\small{$\frac{1}{2}$}}}
\newcommand{\quarter}{\mbox{\small{$\frac{1}{4}$}}}
\newcommand{\threehalves}{\mbox{\small{$\frac{3}{2}$}}}
\newcommand{\la}{\langle}
\newcommand{\ra}{\rangle}
\newcommand{\Nf}{N_{\!f}}

\title{Computation of critical exponent $\eta$ at $O(1/N^3)$ in the four fermi
model in arbitrary dimensions.}
\author{J.A. Gracey, \\ Department of Applied Mathematics and Theoretical
Physics, \\ University of Liverpool, \\ P.O. Box 147, \\ Liverpool, \\
L69 3BX, \\ United Kingdom.}
\date{}
\maketitle
\vspace{5cm}
\noindent
{\bf Abstract.} We solve the conformal bootstrap equations of the four fermi
model or $O(N)$ Gross Neveu model to deduce the fermion anomalous dimension of
the theory at $O(1/N^3)$ in arbitrary dimensions.

\vspace{-16cm}
\hspace{10cm}
{\bf LTH-313}
\newpage
\sect{Introduction.}

The observation that in the neighbourhood of a phase transition in a
(renormalizable) quantum field theory there is a conformal or scaling symmetry
has proved to be a useful technique in solving models within some expansion
scheme, \cite{1,2,3}. For instance, one can determine the critical exponents of
all the Green's functions and these govern the quantum properties of the field
theory. Also, Polyakov demonstrated that the form of the propagators and the
$3$-point vertices are fixed up to constants by demanding that the theory is
symmetric under the conformal group, \cite{4}. Consequently, with this
knowledge various authors formulated a method known as the conformal bootstrap
to solve models such as $\phi^3$ theory in $d$-dimensions order by order in
some perturbative coupling constant, \cite{5,6,7}. The $d$-dimensional
exponents could then be compared with the $\epsilon$-expansion of similar
exponents in other models to ascertain whether they both lay in the same
universality class, \cite{7}. This is important since for physical systems it
is useful to understand which {\em class} of theories underly and describe the
experimental observations and subsequently allows one to make further
predictions from the theoretical model. Following the earlier work of \cite{6}
the ideas surrounding the conformal bootstrap programme were extended to the
$O(N)$ bosonic $\sigma$ model in \cite{8} where $1/N$ replaced the
perturbative coupling constant as the expansion parameter with $N$ large. The
critical exponent $\eta$ of the fundamental field was deduced at $O(1/N^3)$ by
solving a set of bootstrap equations which were derived by using Parisi's
method, \cite{6}, since the $\sigma$ model also involves a $3$-point vertex.
One considers the Dyson equations of the $2$-point function and $3$-vertex
with dressed propagators and vertices. Indeed the $O(1/N^3)$ calculation of
\cite{8} built on earlier $O(1/N^2)$ work of \cite{9,10} where the model was
solved completely at this order by similar conformal techniques. In particular,
the skeleton Dyson equations for the fields were solved by considering, by
contrast, the situation of dressed propagators but undressed vertices. It was
the possibility of solving theories in the large $N$ method this way which
suggested that the full conformal bootstrap could be exploited to probe the
model further.

In this paper, we consider the $O(N)$ Gross Neveu model, \cite{11}, and develop
the analogous formalism to deduce the anomalous dimension, $\eta$, of the
fundamental fermion at $O(1/N^3)$. The viability of such a calculation was also
suggested from the $O(1/N^2)$ work of \cite{12} which used the same techniques
of \cite{9,10} in solving the model at the $d$-dimensional critical point of
the theory defined as the non-trivial zero of the $\beta$-function where one
has a conformal symmetry. Although the $O(N)$ Gross Neveu model describes the
dynamics of the four fermi interaction the fermionic nature of the fields does
not prevent one from using conformal methods and the four point interaction
can be rewritten as a $3$-point vertex plus a bosonic auxiliary field which
means it has the same underlying $\phi^3$ type structure which is
fundamental in this area. Indeed one of the motivations for studying such
models lies in its relation to gauge theories. For instance, if we wish to
probe four dimensional theories with matter fields to high orders in large
$N$ to discover the structure of the quantum theory we must fully comprehend
the way in which fermions have to be dealt with in the formalism. The $O(N)$
Gross Neveu model provides an excellent laboratory for testing such ideas
since it has a $3$-point interaction, in its auxiliary field formulation,
which from the critical point of view mimics the QED interaction. Essentially
one is not obstructed by the tedious complications which occur with the
inclusion of a $\gamma$-matrix at the vertex of integration. Moreover, an
analysis of the four fermi model is necessary and interesting in its own
right. For example, there has been a resurgence of study of the interaction
through its relation to providing an alternative mechanism of mass generation
in the standard model. In \cite{13,14} it was discussed how the effect of the
Higgs boson could be replaced by a composite bosonic field built out of the
binding of two fermions. Further, as our calculations will be in arbitrary
dimensions we will be able to deduce new and useful information on the
structure of the three dimensional model explicitly. This is currently of
interest since four fermi interactions are believed to contribute to models
describing high $T_c$ superconductivity. Thus it is important to have the
basic formalism in place in terms of the values of the underlying three and
four loop Feynman diagrams to ensure that one can eventually extend the
present work to gain predictions for the realistic models which will involve
coupling to a $U(1)$ gauge theory.

The aim therefore of this paper is to provide an extensive analysis of the
four fermi theory at $O(1/N^3)$ in arbitrary dimensions. In \cite{15} the
conformal bootstrap equations which we will solve were derived and checked to
ensure that one could correctly recover the known $O(1/N^2)$ result of
\cite{12}. Its extension to the next order involves the tedious evaluation of
several Feynman graphs. Unlike the $O(N)$ bosonic $\sigma$ model, \cite{8},
there are fewer graphs to consider since graphs with fermion loops with an
odd number of fermions are trivially zero. However, the structure of fermionic
massless Feynman diagrams means that they can in several instances be much
more difficult to compute than their bosonic counterparts.

The paper is organised as follows. In section 2, we present the basic
background to the $O(N)$ Gross Neveu model and introduce the conformal
structure of the fields of the model which are fundamental to the conformal
bootstrap programme. The master equation whose solution will yield the
critical exponent $\eta$ at $O(1/N^3)$ is derived from the three conformal
bootstrap equations of the model in section 3 where we also introduce
Polyakov's conformal triangle approach to formulating the equations. In
section 4, we discuss in detail the techniques required for the evaluation of
the massless graphs contributing to the vertex function which contains
contributions to $\eta$ at $O(1/N^3)$, in a particular limit of the two
regularizing parameters which are present. The main result of the analysis is
presented in section 5 together with several concluding remarks.

\sect{Review of the model.}

In this section we recall previous conformal approaches to solving the $O(N)$
Gross Neveu model in the large $N$ approximation in arbitrary dimensions.
First, we define the lagrangian we use as, \cite{11},
\begin{equation}
L ~=~ \frac{i}{2} \bar{\psi}^i \partialslash \psi^i + \frac{1}{2} \sigma
\bar{\psi}^i \psi^i - \frac{\sigma^2}{2g}
\end{equation}
where $\psi^i$ is the fermion field, $1$ $\leq$ $i$ $\leq$ $N$, and $1/N$ will
be our expansion parameter. The bosonic field $\sigma$ is auxiliary and
eliminating it through its equation of motion yields the four fermi interaction
explicitly. It is more appropriate though to use the formulation (2.1) since
the integration rules to compute massless Feynman integrals are easier to
apply to the situation with $3$-vertices. In the solution of the model in the
usual large $N$ expansion \cite{11} the $\sigma$ field becomes dynamical in
the true vacuum of the quantum theory, though classically it is
non-propagating. The coupling constant, $g$, is dimensionless in two
dimensions, where the model is asymptotically free, \cite{11}. Indeed the
three loop $\beta$-function of (2.1) has been deduced perturbatively in
$\overline{\mbox{MS}}$ using dimensional regularization and in $d$-dimensions
it is, \cite{16,17},
\begin{equation}
\beta(g) ~=~ (d-2)g - (N-2)g^2 + (N-2)g^3 + \quarter(N-2)(N-7)g^4
\end{equation}
where the original two loop calculation was carried out in \cite{18}. The
result (2.2) is what one obtains before setting $d$ $=$ $2$ in minimal
schemes and there is therefore no $d$-dependence in the higher order
coefficients of the coupling constant. Indeed (2.2) is the starting point
for analysing (2.1) in the conformal approach in $d$-dimensions. For $d$
$>$ $2$ one observes that there is a non-trivial zero of the
$\beta$-function at
\begin{equation}
g_c ~ \sim ~ \frac{\epsilon}{(N-2)}
\end{equation}
to one loop where $d$ $=$ $2$ $+$ $\epsilon$. This corresponds to a phase
transition in the theory which has been widely discussed and analysed in three
dimensions in recent years, [19-23].

In statistical mechanics it is well known that in the neighbourhood of a phase
transition physical systems exhibit special properties. For instance, various
(measurable) quantities display a scaling behaviour in that they depend only
on a characteristic length scale such as a correlation length raised to a
certain power known as the critical exponent or index. This exponent totally
characterizes the properties of the critical system. In a completely analogous
way, in the neighbourhood of a phase transition of a (renormalizable) quantum
field theory one observes that there is a conformal symmetry present. In other
words Green's functions are massless and obey a simple power law behaviour.
Further, the exponent of certain Green's functions such as the propagator can
be related to the appropriate critical renormalization group function through a
simple analysis of the renormalization group equation. (See, for example,
\cite{24}.) Indeed the exponent will be a function of the spacetime dimension
$d$ and any other internal parameters of the underlying field theory, which for
(2.1) will be $N$. (At criticality the coupling constant is not independent but
a function of $d$ and $N$.)

Therefore with these simple observations it was possible to solve (2.1) in a
conformal approach which allowed one to go beyond the leading order since the
masslessness of the problem simplifies the intractable integrals, \cite{12},
which would otherwise occur in the conventional large $N$ approach where the
fields are massive. We now recall the essential features which are relevant for
the $O(1/N^3)$ calculation we discuss here. First, since we will now work in
the neighbourhood of $g_c$ given by (2.3) in $d$ $>$ $2$ dimensions we write
down the most general form of the propagators of the theory consistent with
Lorentz and conformal symmetry as, \cite{12},
\begin{equation}
\psi(x) ~ \sim ~ \frac{A\xslash}{(x^2)^\alpha} ~~~,~~~
\sigma(x) ~ \sim ~ \frac{B}{(x^2)^\beta}
\end{equation}
as $x$ $\rightarrow$ $0$ in coordinate space. We have chosen a propagating
$\sigma$ field since we are solving the theory in the true vacuum. Whilst it
is more convenient to compute in $x$-space one can easily deduce the
momentum space forms of (2.4) through the Fourier transform
\begin{equation}
\frac{1}{(x^2)^\alpha} ~=~ \frac{a(\alpha)}{\pi^\mu 2^{2\alpha}}
\int_k \frac{e^{ikx}}{(k^2)^{\mu-\alpha}}
\end{equation}
valid for all $\alpha$ where we set the spacetime dimension $d$ to be $d$ $=$
$2\mu$ for later convenience. The quantity $a(\alpha)$ is defined by
$a(\alpha)$ $=$ $\Gamma(\mu-\alpha)/\Gamma(\alpha)$. In (2.4) the quantities
$A$ and $B$ are the amplitudes of the respective fields and are
$x$-independent, whilst we have defined the dimensions of the fields to be
$\alpha$ and $\beta$. Their canonical dimensions can be determined by a
dimensional analysis of the action which is dimensionless. However, quantum
mechanically these dimensions do not remain as their engineering values due to
quantum fluctuations such as radiative corrections. To allow for this scenario
we define, \cite{12},
\begin{equation}
\alpha ~=~ \mu + \half \eta ~~~,~~~ \beta ~=~ 1 - \eta - 2 \Delta
\end{equation}
where $\eta$ is the anomalous dimension of the fermion and from an examination
of the $3$-vertex of (2.1), $2\Delta$ is its anomalous dimension. Both $\eta$
and $\Delta$ are $O(1/N)$ and depend on $d$ and $N$. In a critical point
analysis of the renormalization group function, $\eta$ corresponds to the wave
function renormalization and $\eta$ $+$ $2\Delta$ to the mass anomalous
dimension. They have both been determined at $O(1/N^2)$ in arbitrary dimensions
as, \cite{12,25},
\begin{eqnarray}
\eta_1 &=& \frac{2\Gamma(2\mu-1)(\mu-1)^2}{\Gamma(2-\mu)\Gamma(\mu+1)
\Gamma^2(\mu)} \\
\Delta_1 &=& \frac{\mu \eta_1}{2(\mu-1)}
\end{eqnarray}
\begin{equation}
\eta_2 ~=~ \frac{\eta^2_1}{2(\mu-1)^2} \left[ \frac{(\mu-1)^2}{\mu} + 3\mu
+ 4(\mu-1) + 2(\mu-1)(2\mu-1)\Psi(\mu) \right]
\end{equation}
\begin{equation}
\Delta_2 ~=~ \frac{\mu \eta^2_1}{2(\mu-1)^2} \left[ \frac{}{} 3\mu(\mu-1)
\Theta + (2\mu-1)\Psi - \frac{(2\mu-1)(\mu^2-\mu-1)}{(\mu-1)} \right]
\end{equation}
where $\eta$ $=$ $\sum_{i=1}^\infty \eta_i/N^i$, $\Delta$ $=$
$\sum_{i=1}^\infty \Delta_i/N^i$, $\Psi(\mu)$ $=$ $\psi(2\mu-1)$ $-$ $\psi(1)$
$+$ $\psi(2-\mu)$ $-$ $\psi(\mu)$, $\Theta(\mu)$ $=$ $\psi^\prime(\mu)$ $-$
$\psi^\prime(1)$ and $\psi(x)$ is the logarithmic derivative of the
$\Gamma$-function. The result (2.7) has been given originally in \cite{26}
whilst (2.8) has been determined independently in \cite{21}. It is the aim of
this paper to deduce $\eta_3$ in arbitrary dimensions. It is worth noting that
the $O(1/N^2)$ correction to the critical exponent $\lambda$ which relates to
the $\beta$-function of (2.1) has recently been determined too as,
\cite{12,27},
\begin{equation}
\lambda_1 ~=~ - \, (2\mu-1)\eta_1
\end{equation}
\begin{eqnarray}
\lambda_2 &=& \frac{2\mu\eta^2_1}{(\mu-1)} \left[ \frac{2}{(\mu-2)^2\eta_1}
- \frac{(2\mu-3)\mu}{(\mu-2)}(\Phi + \Psi^2) \right. \nonumber \\
&+& \left. \Psi \left( \frac{1}{(\mu-2)^2} + \frac{1}{2(\mu-2)} - 2 \mu^2
- \frac{3}{2} - \frac{1}{2\mu} - \frac{3}{(\mu-1)} \right) \right. \nonumber \\
&+& \left. \frac{3\mu\Theta}{4} \left( 9 - 2\mu + \frac{6}{(\mu-2)} \right)
+ 2\mu^2 - 5\mu - 3 + \frac{5}{4\mu} \right. \nonumber \\
&-& \left. \frac{1}{4\mu^2} - \frac{7}{2(\mu-1)} - \frac{1}{(\mu-1)^2}
+ \frac{1}{4(\mu-2)} - \frac{1}{2(\mu-2)^2} \right]
\end{eqnarray}
where $\Phi$ $=$ $\psi^\prime(2\mu-1)$ $-$ $\psi^\prime(2-\mu)$ $-$
$\psi^\prime(\mu)$ $+$ $\psi^\prime(1)$ which means (2.1) has been solved
completely at $O(1/N^2)$. It is important to recognise that these analytic
expressions have been deduced merely from the ans\"{a}tze (2.4) and the
solution of the critical Schwinger Dyson equations, which therefore means
that the method of \cite{9,10,12} is an extremely powerful one.

With (2.4) it is possible to deduce the scaling behaviour of the respective
$2$-point functions which we will require here. Their coordinate space forms
are deduced by first mapping (2.4) to momentum space via (2.5), inverting
the propagator before applying the inverse map to coordinate space,
\cite{9}. Thus, we have, \cite{12}
\begin{equation}
\psi^{-1}(x) ~ \sim ~ \frac{r(\alpha-1)\xslash}{A(x^2)^{2\mu-\alpha}} ~~~,~~~
\sigma^{-1}(x) ~ \sim ~ \frac{p(\beta)}{B(x^2)^{2\mu-\beta}}
\end{equation}
where the functions have an analogous scaling structure but the amplitudes
are non-trivially related. The functions $r(\alpha-1)$ and $p(\beta)$ are
defined as
\begin{equation}
p(\beta) ~=~ \frac{a(\beta-\mu)}{\pi^{2\mu}a(\beta)} ~~~,~~~
r(\alpha) ~=~ \frac{\alpha p(\alpha)}{(\mu-\alpha)}
\end{equation}
It is the quantities (2.13) and (2.14) which will be necessary to solve the
conformal bootstrap equations.

\sect{Conformal bootstrap equations.}

In this section, we recall the basic bootstrap equations for the model which
will then be solved at $O(1/N^2)$ for $\eta$ and which were given in \cite{15}
for (2.1). The development of a bootstrap programme to solving field theories
dates back to the early work of \cite{4,6}. Basically one simplifies the set
of Feynman diagrams contributing to a Green's function by first using
dressed propagators. Further, each vertex of the Feynman graph is replaced by
a conformal triangle which was discussed originally in \cite{4} and later in
\cite{8}. To understand this concept we first have to introduce the
technique known as uniqueness which was first discussed in \cite{28} and is
a fundamental tool for evaluating massless Feynman diagrams. The rule for a
purely bosonic vertex was given in \cite{28,10} whilst the development to the
style of vertex which arises in the model (2.1) was given in \cite{12} and is
illustrated in fig. 1. When the arbitrary exponents $\beta_i$ are constrained
to be their uniqueness value $\sum_{i=1}^3 \beta_i$ $=$ $2\mu$ $+$ $1$ then
one can compute the integral over the vertex of integration of the left side
of fig. 1 to obtain the product of three propagators where $\nu(\beta_1,
\beta_2, \beta_3)$ $=$ $\pi^\mu \prod_{i=1}^3 a(\beta_i)$. The origin of this
uniqueness condition can be understood in several ways. First, if one carries
out the explicit integral using Feynman parameters over the vertex of
integration one finds that for general $\beta_i$ the calculation cannot be
completed due to the appearance of a hypergeometric function. The analysis can
only proceed if one chooses the sum of the exponents to be constrained to
$2\mu$ $+$ $1$, whence the hypergeometric function becomes a simple
algebraic function. Alternatively one can make a conformal transformation on
the integral representing the graph of the left side of fig. 1. If $z$ is the
location of the integration vertex and the origin is chosen to be at the
external end of the bosonic line then one can make the following change of
variables, \cite{15},
\begin{equation}
z_\mu ~ \longrightarrow ~ \frac{z_\mu}{z^2}
\end{equation}
which represents a conformal transformation. The part of the numerator
involving fermion propagators transforms as, say,
\begin{equation}
(\xslash-\zslash) ~ \longrightarrow ~ - \, \frac{\xslash(\xslash
-\zslash)\zslash}{x^2 z^2}
\end{equation}
This results in a new integral of the same structure but where the exponents
of the lines are related to the exponents of the original graph. In
particular the exponent of the bosonic line becomes $2\mu$ $+$ $1$ $-$
$\sum_i\beta_i$. Therefore choosing this to be zero one obtains a simple
chain integral and the result of fig. 1 emerges naturally. We have detailed
this type of conformal change of variables (3.1) and (3.2) here since it
is fundamental to the problem and is the starting technique to the
computation of all our graphs.

With this basic rule one can define the conformal triangle construction for the
vertex of the model. It is illustrated in fig. 2 for the $\sigma \bar{\psi}
\psi$ vertex of (2.1) where the internal exponents $a$ and $b$ are defined to
be such that each vertex of the triangle is unique ie
\begin{equation}
2a+\beta ~=~ a + \alpha + b ~=~ 2 \mu + 1
\end{equation}
It is worth noting that the way the model is defined $2\alpha$ $+$ $\beta$ $=$
$2\mu$ $+$ $1$ $-$ $2\Delta$ so that the vertex of (2.1) is only unique at
leading order in $1/N$, since $\Delta$ $=$ $O(1/N)$. The idea then of
\cite{6,8} is to replace all vertices, which are therefore not conformal, by
a triangle which after either conformal transformations or applying the rule
of fig. 1 yields the original vertex, up to a (amplitude) factor. In the
conformal bootstrap approach all vertices of a Feynman diagram are replaced
by conformal triangles. In other words each vertex of this new graph is unique
or conformal and it is therefore elementary to observe that such graphs are
then proportional to the basic (bare) graph by using conformal transformations
which in turn allows one to write down simple equations for the Dyson
equations of the Green's functions.

For example, the $3$-point vertex of (2.1) is given by an infinite sum of
graphs, where we will use $1/N$ as our ordering parameter, and we have
illustrated the first few graphs in its expansion in fig. 3 where the labels
$\Gamma_i$ will be used for later purposes. Each vertex of $\Gamma_i$ and
the left side are replaced by the construction of fig. 2 and the values of
the integral they correspond to will be a function of the exponents $\alpha$
and $\beta$ and the quantity $z$ defined as $z$ $=$ $fA^2B$ where $A$ and $B$
are our earlier amplitudes and $f$ is the amplitude of the triangle. These
are the three basic variables of the formalism and to deduce information on
each we need three bootstrap equations. The vertex expansion of fig. 3
provides the first. Since each graph of fig. 3 will be a function of $\alpha$,
$\beta$ and $z$ and also by our arguments be proportional to the basic
vertex, then if we denote by $V$ the sum of all contributions from the
graphs of fig. 3 it must be unity. Thus the first bootstrap equation is,
\cite{8,15},
\begin{equation}
V(z,\alpha,\beta;0,0) ~=~ 1
\end{equation}
where we have displayed the arguments of the vertex function explicitly. The
origin of the final two arguments will become apparent later when we have to
introduce two regularizing parameters. Thus (3.4) represents the sum of all
the contributions of the graphs in fig. 3.

The remaining two bootstrap equations have been derived in \cite{15} by
following the analogous construction of \cite{6} for a $\phi^3$ theory in
arbitrary dimensions. Rather than repeat that derivation we will record their
form and then make several comments. We have, \cite{15},
\begin{eqnarray}
r(\alpha-1) &=& zt \, \left. \frac{\partial ~}{\partial \epsilon^\prime}
V(z,\alpha,\beta;\epsilon,\epsilon^\prime) \right|_{\epsilon \, = \,
\epsilon^\prime \, = \, 0} \\
\frac{p(\beta)}{N} &=& zt \, \left. \frac{\partial ~}{\partial \epsilon}
V(z,\alpha,\beta;\epsilon,\epsilon^\prime) \right|_{\epsilon \, = \,
\epsilon^\prime \, = \, 0}
\end{eqnarray}
where
\begin{equation}
t ~=~ \frac{\pi^{4\mu} a^2(\alpha-1) a^2(a-1) a(b) a(\beta)}
{\Gamma(\mu) (\alpha-1)^2 (a-1)^2 a(\beta-b)}
\end{equation}
The quantities $\epsilon$ and $\epsilon^\prime$ are infinitesimal regularizing
parameters introduced into the formalism by shifting $\alpha$ and $\beta$
respectively by $\alpha$ $\rightarrow$ $\alpha$ $+$ $2\epsilon^\prime$, $\beta$
$\rightarrow$ $\beta$ $+$ $2\epsilon$. Their introduction is necessary to avoid
ill defined quantities in the derivation of the conformal bootstrap equations
of the $2$-point functions of (3.5) and (3.6). Basically a divergent integral
arises which needs to be regularized and this is achieved by the above shift
though the integral appears multiplied by either $\epsilon$ or
$\epsilon^\prime$ so that only the residue of the simple pole, ie $t$, is
required.

The effect of the introduction of the regularization is that the sum of graphs
contributing to the vertex function $V$ become additionally functions of
$\epsilon$ and $\epsilon^\prime$, ie $V(\alpha,\beta,z;\epsilon,
\epsilon^\prime)$. The graphs remain conformal because the internal exponents
of the conformal triangle are adjusted (symmetrically) to preserve the
uniqueness of each vertex. Thus with the regulators the vertex function is
a sum of values which now depend on $z$, $\alpha$, $\beta$, $\epsilon$ and
$\epsilon^\prime$ and it is this function which appears in (3.4)-(3.6).
Consequently, one has to compute the graphs of fig. 3 in the presence of the
shift.

With (3.5) and (3.6), it is now possible to deduce a master equation which
determines $\eta_3$. First, we let $V_1$ denote the contribution to the
vertex function from the one loop graph of fig. 3 and $V_2$ the higher order
pieces which will be $O(1/N)$ relative to $V_1$. The former will later be
expanded in powers of $1/N$ too. Then taking the quotient of (3.5) and (3.6)
\begin{equation}
\frac{Nr(\alpha-1)}{p(\beta)} ~=~ \left. \frac{\partial V}{\partial
\epsilon^\prime}\right| \bigg/ \left. \frac{\partial V}{\partial \epsilon}
\right|
\end{equation}
On the left side of (3.8) we have a function of the exponents which can be
expanded in powers of $1/N$ and will involve the anomalous exponents $\eta$ and
$2\Delta$. Expanding to $O(1/N^2)$ the only unknown there is is $\eta_3$ since
$\eta_1$, $\eta_2$, $\Delta_1$ and $\Delta_2$ have already been determined.
Thus to deduce $\eta_3$ the right side of (3.8) must be expanded to $O(1/N^2)$
and this will therefore involve the set of graphs contributing to $V_2$.

However, it is worth remarking on the functional structure of the graphs
themselves. In representing each vertex by a conformal triangle one is
replacing a vertex of the theory which is non-unique by a construction where
each vertex is unique. The non-uniqueness of this is reflected in the form
of the value of the conformal triangle, which can easily be deduced by the
method of subtractions discussed in \cite{10,12}. In particular it will be
of the form $1/\Delta$ where the residue is not important for the moment,
\cite{29}. In other words the variation from the value of the exponents which
make the vertex conformal or unique appears as a pole in the value of the
vertex. Therefore, a graph built out of $n$ conformal triangles will have the
structure $h(\Delta)/\Delta^n$ where $h(\Delta)$ is an analytic function of
$\Delta$. In the regularized theory the variation from the conformal value is
$(\Delta$ $-$ $\epsilon)$ and $(\Delta$ $-$ $\epsilon^\prime)$ in the vertices
which are regularized. Therefore the form of a graph contributing to the vertex
functions $V_1$ and $V_2$ will become
$h(\Delta,\epsilon,\epsilon^\prime)/[\Delta^{n-2}(\Delta-\epsilon)
(\Delta-\epsilon^\prime)]$. It is this structure which we now consider in the
context of (3.8). Differentiating with respect to either regularizing
parameters involves differentiating the residue or the simple pole in
$(\Delta-\epsilon)$ or $(\Delta-\epsilon^\prime)$. In the case of the latter
the sum of all such contributions will be $V/(\Delta-\epsilon)|$ or
$V/(\Delta-\epsilon^\prime)|$, whilst in the former case it will be the sum of
the residue contributions. The upshot of this is that within (3.8) when
$\epsilon$ and $\epsilon^\prime$ are set to zero the contribution from the
poles cancels and only the residue needs to be considered. In other words,
\begin{equation}
\frac{Nr(\alpha-1)}{p(\beta)} ~=~ \left[ 1 + \Delta \left. \frac{\partial
V}{\partial \epsilon^\prime} \right|_{\mbox{res}} \right]\left[ 1 +
\Delta \left. \frac{\partial V}{\partial \epsilon} \right|_{\mbox{res}}
\right]^{-1}
\end{equation}
exactly where we have used (3.4) and the subscript $\mbox{res}$ denotes
differentiation with respect to $\epsilon$ or $\epsilon^\prime$ of the residue
of the vertex function. Therefore it is the expansion of each of these
quantities to $O(1/N^2)$ which is required for $\eta_3$. More concretely we
have
\begin{eqnarray}
\frac{Nr(\alpha-1)}{p(\beta)} &=& 1 + \frac{\Delta_1}{N} \left(
\left. \frac{\partial V_{10}}{\partial \epsilon^\prime} \right|
- \left. \frac{\partial V_{10}}{\partial \epsilon} \right| \right)
+ \frac{1}{N^2} \left[ \Delta_2 \left( \left. \frac{\partial V_{10}}
{\partial \epsilon^\prime} \right|
- \left. \frac{\partial V_{10}}{\partial \epsilon} \right| \right)
\right. \nonumber \\
&+& \left. \Delta_1 \left( \left. \frac{\partial V_{11}}
{\partial \epsilon^\prime} \right|
- \left. \frac{\partial V_{11}}{\partial \epsilon} \right| \right)
- \Delta^2_1 \left. \frac{\partial V_{10}}{\partial \epsilon} \right|
\left. \left( \frac{\partial V_{10}}{\partial \epsilon^\prime} \right|
- \left. \frac{\partial V_{10}}{\partial \epsilon} \right| \right)
\right. \nonumber \\
&+& \left. \Delta_1 \left( \left. \frac{\partial V_2}{\partial
\epsilon^\prime}\right|
- \left. \frac{\partial V_2}{\partial \epsilon}\right|\right) \right]
\end{eqnarray}
where $V_1$ $=$ $V_{10}$ $+$ $V_{11}/N$ and we now omit any comment on the
evaluation symbol and $z_2$ is deduced from (3.4). Thus (3.10) forms the
master equation to deduce $\eta_3$. It has a two part structure. First, one
requires the leading order graph $V_1$ to be expanded to $O(1/N^2)$ and to be
computed for non-zero $\epsilon$ and $\epsilon^\prime$, whilst the graphs
for $V_2$ must be calculated. In \cite{15} which established the bootstrap
equations the vertex function for $V_1$ was determined in order to check that
the already known $O(1/N^2)$ results of \cite{12} could be correctly
determined. Moreover, the function was given at $O(1/N^2)$ for non-zero
$\epsilon$ and $\epsilon^\prime$ and we recall
\begin{equation}
\Gamma_1 ~=~ - \, \frac{Q^3}{\Delta(\Delta-\epsilon)(\Delta-\epsilon^\prime)}
\exp [ F(\epsilon, \epsilon^\prime,\Delta)]
\end{equation}
where
\begin{eqnarray}
F(\epsilon,\epsilon^\prime,\Delta) &=& \left( 5B_\beta - 2B_{\alpha-1}
-3B_0 - \frac{2}{\alpha-1}\right)\Delta - (B_\beta - B_0) \epsilon \nonumber \\
&+& \left( B_0 - B_{\alpha-1} - \frac{1}{\alpha-1} \right) \epsilon^\prime
\nonumber \\
&+& \left( C_{\alpha-1} - \frac{7C_\beta}{2} - \frac{3C_0}{2}
- \frac{1}{(\alpha-1)^2} \right)\Delta^2 \nonumber \\
&+& \left( C_\beta + C_0 - 2C_{\alpha-1} + \frac{2}{(\alpha-1)^2} \right)
\Delta\epsilon \nonumber \\
&+& \left( C_0 - C_\beta - 2C_{\alpha-1} + \frac{2}{(\alpha-1)^2} \right)
\Delta \epsilon^\prime
\end{eqnarray}
with $B_x$ $=$ $\psi(\mu-x)$ $+$ $\psi(x)$, $B_0$ $=$ $\psi(1)$ $+$
$\psi(\mu)$, $C_x$ $=$ $\psi^\prime(x)$ $-$ $\psi^\prime(\mu-x)$, $C_0$ $=$
$\psi^\prime(\mu)$ $-$ $\psi^\prime(1)$ and
\begin{equation}
Q ~=~ - \, \frac{\pi^{2\mu} a^2(\alpha-1) a(\beta)}{(\alpha-1)^2\Gamma(\mu)}
\end{equation}
We have illustrated the full $\Gamma_1$ graph in terms of conformal triangles
in fig. 4 to show where the regulators appear in the appropriate exponents
which is important for writing down the graphs of $V_2$. It is crucial to note
that in (3.12) one should substitute $\alpha$ $=$ $\mu$ $+$ $\half \eta$ and
$\beta$ $=$ $1$ $-$ $\eta$ since the calculation of the graph was simplified by
having the exponent $\Delta$ of the $\sigma$ line displayed explicitly. With
(3.12) it is a trivial matter to deduce $\eta_2$ agrees with (2.9) where the
variable $z$ is eliminated through
\begin{equation}
1 ~=~ - \, \frac{z Q^3}{\Delta^3}
\end{equation}
at leading order from (3.4).

\sect{Computation of higher order graphs.}

All that remains is the evaluation of the higher order graphs which is far
from a trivial exercise. The basic tools to compute the graphs are the
conformal transformations (3.1) and (3.2) on each of the vertices of
integration of the graph with the vertex joining to the external $\sigma$
line chosen to be the origin and the uniqueness rule of fig. 1 and its
bosonic counterpart, \cite{28,10}. Whilst it is possible to compute each
graph with these methods easily when $\epsilon$ $=$ $\epsilon^\prime$ $=$ $0$
which is required for the subsequent correction to (3.14), it turned out that
it was not possible to evaluate the graphs exactly when $\epsilon$ $\neq$
$0$, $\epsilon^\prime$ $\neq$ $0$. It transpired, however, that the
difference, ie
\begin{equation}
\left. \frac{\partial \Gamma_i}{\partial \epsilon^\prime} \right|
- \left. \frac{\partial \Gamma_i}{\partial \epsilon} \right|
\end{equation}
could be determined in each case, $i$ $=$ $2, \ldots, 5$. There are four higher
order graphs to consider. Whilst $\Gamma_4$ and $\Gamma_5$ are equivalent in
the absence of the regulators, in the regularized version they are not equal
since the bottom right external vertex of each $\Gamma_i$ is regularized. In
the evaluation of $\Gamma_i$ we replace each unregularized conformal triangle
with the original vertex and multiply the graph by $-\, Q/\Delta$ from (3.11),
for each such vertex. It is only the regularized vertices which will give the
contributions to the residue of $V_2$ which are relevant for $\eta_3$.

Before discussing the determination of $\Gamma_3$, $\Gamma_4$ and $\Gamma_5$ we
detail the calculation of $\Gamma_2$ as it will illustrate some elementary
steps which occur in all cases. Making a conformal transformation on $\Gamma_2$
in the manner indicated yields a three loop integral which is illustrated in
fig. 5 and we have evaluated two elementary chain integrals en route whose
integration rule has been given in \cite{12}, for example. However, this
integral has two unique vertices which leads to the $2$-loop integral of fig. 5
where we have now set $\Delta$ $=$ $0$ since it does not contribute to the
residue of $\Gamma_2$ at this order in $1/N$. (The poles in $\Delta$ emerge
from factors such as $a(\mu-\Delta)$ which arise from a chain integral and the
other from the integration of the two unique vertices.) In the notation of
\cite{27} it is $\la \tilde{\alpha}, \beta-\epsilon^\prime, \widetilde{\alpha
-\epsilon^\prime}, \beta, \epsilon+\epsilon^\prime \ra$ but it cannot be
evaluated by uniqueness methods for arbitrary $\epsilon$ and $\epsilon^\prime$.
However, if we recall that it is the difference of the derivatives of the
residues which is relevant for $\eta_3$ we can make use of the observation that
\begin{eqnarray}
&& \left. \left( \frac{\partial ~}{\partial \epsilon^\prime} -
\frac{\partial ~ }{\partial \epsilon} \right)
\la \tilde{\alpha}, \beta-\epsilon^\prime, \widetilde{\alpha-\epsilon^\prime},
\beta, \epsilon+\epsilon^\prime \ra \right|_{\epsilon \, = \,
\epsilon^\prime \, = \, 0} \nonumber \\
&& = ~ \frac{\partial ~ }{\partial \epsilon^\prime} \left.
\la \tilde{\alpha}, \beta-\epsilon^\prime, \widetilde{\alpha-\epsilon^\prime},
\beta, 0 \ra \right|_{\epsilon^\prime \, = \, 0}
\end{eqnarray}
to evaluate the relevant piece. The latter integral is equivalent to chain
integrals and is simply
\begin{equation}
\frac{\nu^2(\alpha-1,\alpha-1,\beta)}{(\alpha-1)^4} \left[ B_\beta
- B_{\alpha-1} - \frac{1}{(\alpha-1)} \right]
\end{equation}
The overall value of the contribution is then given by carrying out the
differentiation with respect to the regularizing parameters on the functions
obtained after the successive integrations, which is elementary. We find
\begin{equation}
\left. \left( \frac{\partial\Gamma_2}{\partial \epsilon^\prime}
- \frac{\partial \Gamma_2}{\partial \epsilon} \right) \right|
{}~=~ - \, \frac{2Q^5\nu^2(\alpha-1,\alpha-1,\beta)}{(\alpha-1)^4\Delta^5}
\left[ B_\beta - B_{\alpha-1} - \frac{1}{(\alpha-1)} \right]
\end{equation}
Including the amplitude factor $z^2$ associated with the graph and using (3.14)
we have
\begin{equation}
\left. \left( \frac{\partial V_{2,2}}{\partial \epsilon^\prime}
- \frac{\partial V_{2,2}}{\partial \epsilon} \right) \right|
{}~=~ - \, \frac{\mu\eta_1}{(\mu-1)^3N}
\end{equation}
in an obvious notation where we have substituted for the leading order values
of $\alpha$ and $\beta$ with respect to $1/N$.

The treatment of $\Gamma_3$ is equally as straightforward as that of $\Gamma_2$
and we note that
\begin{eqnarray}
\left. \left( \frac{\partial\Gamma_3}{\partial \epsilon^\prime}
- \frac{\partial \Gamma_3}{\partial \epsilon} \right) \right|
&=& - \, \frac{\pi^{4\mu} Q^7 a(2\mu-2)a(\mu-1)}{\Delta^7(\mu-1)^6}
\left[ (\mu-1)(2\mu-1) \right. \nonumber \\
&-& \left. (2\mu^2-5\mu+4)\Psi - \frac{5(2\mu^2-5\mu+4)}{2(\mu-1)} \right]
\end{eqnarray}
from which we obtain
\begin{eqnarray}
\left. \left( \frac{\partial V_{2,3}}{\partial \epsilon^\prime}
- \frac{\partial V_{2,3}}{\partial \epsilon} \right) \right|
&=& - \, \frac{\mu\eta_1}{(\mu-1)^2N} \left[ (\mu-1)(2\mu-1) \right. \nonumber
\\
&-& \left. (2\mu^2-5\mu+4)\Psi - \frac{5(2\mu^2-5\mu+4)}{2(\mu-1)} \right]
\end{eqnarray}
This calculation made use of the elementary chain integral
\begin{eqnarray}
\int_y \frac{(-\yslash)(\yslash-\xslash)\mbox{tr}(-\yslash)(\yslash
-\xslash)}{(y^2)^\alpha ((x-y)^2)^\beta}
&=& \frac{[2(\mu-\alpha)(\mu-\beta) + 3\mu - \alpha - \beta]}{(\alpha-1)
(\beta-1)} \nonumber \\
&\times& \nu (\alpha-1, \beta-1, 2\mu-\alpha-\beta+2)
\end{eqnarray}

The analysis of both $\Gamma_4$ and $\Gamma_5$ turned out to be extremely
intricate and involved and it is worth discussing several of the intermediate
steps in one due to the presence of a common difficult integral in each. We
consider $\Gamma_4$ which after various transformations and elementary
integrals leads us to consider the derivative with respect to $\delta$ of the
two loop integral of fig. 6 where there is a trace over the lower right and
one of the top fermion propagators. This is, in fact, a remnant of the
fermion loop of the original graph. To evaluate this graph to $O(\delta)$ we
restricted attention to the leading order values since that leads to an
integral where most of the exponents are unity which is convenient for
taking a fermion trace. For instance, taking the trace over the open fermion
propagators and dividing by the appropriate factor $\mbox{tr} 1$, fig. 6 is
equivalent to
\begin{equation}
\half \mbox{tr} [ \la 1, \tilde{2}, \tilde{1}, 1-\delta,\mu-1 \ra
- \la 0, \tilde{2}, \tilde{1}, 1-\delta,\mu-1 \ra
- \la 1, \tilde{1}, \tilde{1}, 1-\delta,\mu-1 \ra ]
\end{equation}
The second term is trivial to deduce whilst the third becomes trivial after
making the transformation $\leftarrow$ in the notation of \cite{10}. The
hardest part of the integral lurks within the first term, which after
performing the trace explicitly yields one trivial graph and
\begin{equation}
\la 1,2,1,1-\delta,\mu-2\ra ~-~ \la 1,1,1,1-\delta,\mu-1\ra
\end{equation}
The $\delta$ expansion of an integral similar to the second term of (4.10)
arises in the analogous $O(1/N^3)$ calculation in the pioneering work of
\cite{8}. Unfortunately it is not possible to give a closed form for the
$O(\delta)$ correction in terms of elementary functions like $\Psi$ and
$\Theta$ for all dimensions. Instead one is forced to leave the $O(\delta)$
correction defined as the quantity $I(\mu)$. Thus
\begin{equation}
\la \mu-1,1,1-\delta,\mu-1,\mu-1+\delta\ra ~=~ ChT(1,1) [ 1 + \delta I(\mu)
+ O(\delta^2) ]
\end{equation}
where we have used the result $ChT(1,1-\delta)$ $=$ $ChT(1,1+\delta)$ and the
function $ChT(\alpha,\beta)$ is defined to be $\la \alpha,\mu-1,\mu-1,\beta,
\mu-1\ra$ for all $\alpha$ and $\beta$ in the notation of \cite{10} where it
was evaluated exactly as \cite{10}
\begin{eqnarray}
ChT(\alpha,\beta) &=& \frac{\pi^{2\mu}a(2\mu-2)}{\Gamma(\mu-1)} \left[
\frac{a(\alpha)a(2-\alpha)}{(1-\beta)(\alpha+\beta-2)}
+ \frac{a(\beta)a(2-\beta)}{(1-\alpha)(\alpha+\beta-2)} \right. \nonumber \\
&+& \left. \frac{a(\alpha+\beta-1)a(3-\alpha-\beta)}{(\alpha-1)(\beta-1)}
\right]
\end{eqnarray}
Whilst a closed form for $I(\mu)$ is not available it can be analysed using
the Gegenbauer polynomial techniques of \cite{30} to obtain a set of double
sums over $\Gamma$-functions and its $\epsilon$-expansion near $d$ $=$ $2$
$+$ $\epsilon$ can be given to several orders. Then with the definition
\begin{equation}
I(\mu) ~=~ - \, \frac{2}{3(\mu-1)} ~+~ \Xi(\mu)
\end{equation}
we have
\begin{equation}
\Xi(\mu) ~=~ \frac{\zeta(3)\epsilon^2}{6} - \frac{\zeta(4)\epsilon^3}{8}
+ \frac{13\zeta(5)\epsilon^4}{48} + O(\epsilon^5)
\end{equation}
where the first few terms of (4.14) were given in \cite{10} and later ones in
\cite{31,32}. It turns out that in three dimensions an exact expression can
be deduced as \cite{10},
\begin{equation}
I(\threehalves) ~=~ 2 \ln 2 + \frac{3\psi^{\prime\prime}(\half)}{2\pi^2}
\end{equation}
The remaining integral of (4.10) also contains $I(\mu)$ which can be deduced
by various transformations of bosonic two loop integrals given in \cite{10} and
recursion relations of \cite{33}. One result which was required in this and
which is worth recording is
\begin{eqnarray}
&&\left. \frac{\partial ~}{\partial \epsilon} \la \mu-1+\epsilon,2,1,\mu-1,
\mu-2\ra \right| \nonumber \\
\!\!&=& \pi^{2\mu} a(1)a(2\mu-2)
\left[ \frac{(2\mu-3)(\mu-3)}{(\mu-2)^2} - \frac{2(\mu-1)}{(\mu-2)}
+ \frac{(2\mu-3)\Psi}{(\mu-2)} \right. \nonumber \\
&+& \!\!\!\left. \frac{(2\mu-3)}{2(\mu-1)(\mu-2)} - \mu \left( \Theta +
\frac{1}
{(\mu-1)^2} \right) - 3(\mu-2)I(\mu) \left( \Theta + \frac{1}{(\mu-1)^2}
\right) \right] \nonumber \\
\end{eqnarray}
Consequently we have that the coefficient of the $O(\delta)$ term of
(4.9) is
\begin{eqnarray}
&-& \frac{\pi^{2\mu}a(1)a(2\mu-2)}{2(\mu-1)} \left[ 3(\mu-1)\Theta \Xi
+ \frac{3\Xi}{(\mu-1)} \right. \nonumber \\
&&+ \left. (\mu-5)\Theta + \frac{2\mu\Psi}{(\mu-1)} + \frac{(2\mu-3)}
{(\mu-1)^2} \right]
\end{eqnarray}
Collecting all the contributions to the integral and evaluating the
derivative of the residue gives the relatively simple result
\begin{eqnarray}
\left. \left( \frac{\partial V_{2,4}}{\partial \epsilon^\prime}
- \frac{\partial V_{2,4}}{\partial \epsilon} \right) \right|
&=& \frac{\mu}{N} \left[ \frac{3\Theta\Xi}{2} + \frac{3\Xi}{2(\mu-1)^2}
+ \frac{\mu\Psi}{(\mu-1)^2} \right. \nonumber \\
&&+ \left. \frac{(\mu-8)\Theta}{2(\mu-1)} + \frac{(2\mu-3)}{(\mu-1)^3} \right]
\end{eqnarray}

The procedure for $\Gamma_5$ is equally as tedious but does not merit
extensive coverage since the key techniques have already been covered in the
discussion of $\Gamma_4$. A similar integral to fig. 6 occurs but this time
the regulator is not on the line with exponent $\beta$ but on the other
bosonic line appearing as $(\mu-\beta-\epsilon)$. Its $O(\epsilon)$
contribution is
\begin{equation}
- \, \frac{\pi^{2\mu}a(1)a(2\mu-2)}{2(\mu-1)} \left[ 3(\mu-1)\Theta\Xi
+ \frac{3\Xi}{(\mu-1)} - (3\mu+2)\Theta + \frac{2\mu\Psi}{(\mu-1)} \right]
\end{equation}
as an intermediate check on the calculation. Overall we find
\begin{eqnarray}
\left. \left( \frac{\partial V_{2,5}}{\partial \epsilon^\prime}
- \frac{\partial V_{2,5}}{\partial \epsilon} \right) \right|
&=& \frac{\mu}{N} \left[ \frac{3\Theta\Xi}{2} + \frac{3\Xi}{2(\mu-1)^2}
+ \frac{\mu\Psi}{(\mu-1)^2} \right. \nonumber \\
&-& \left. \frac{4\Theta}{(\mu-1)} - \frac{(2\mu-5)(\mu-2)}{2(\mu-1)^3} \right]
\end{eqnarray}
This completes the discussion of the contributions to $V_2$. However, it is
worth recording the values of each of the $\Gamma_i$ when $\epsilon$ $=$
$\epsilon^\prime$ $=$ $0$ since they are required for the higher order
corrections to (3.14) which determines $z_2$. We found that
\begin{equation}
\Gamma_2 ~=~ - \, \frac{Q^5}{\Delta^5} \frac{\nu^2(\alpha-1,\alpha-1,\beta)}
{(\alpha-1)^4}
\end{equation}
\begin{equation}
\Gamma_3 ~=~ - \, \frac{Q^7}{\Delta^7} \frac{\nu^3(\alpha-1,\alpha-1,
\beta)}{(\alpha-1)^6\beta^2} \nu(\beta,\beta,2\mu-2\beta)
[2(\mu-\beta)(\mu-\beta-1)+\mu]
\end{equation}
\begin{equation}
\Gamma_4 ~=~ \Gamma_5 ~=~ \frac{\pi^{2\mu}a(1)a(2\mu-2) Q^7\nu^2(\alpha-1,
\alpha-1,\beta)}{2\Delta^7(\alpha-1)^4} \left[ 3\Theta - \frac{(2\mu-3)}
{(\mu-1)^2} \right]
\end{equation}

\sect{Discussion.}

With the explicit values of the $V_2$ contributions to (3.10) calculated it
is now possible to deduce $\eta_3$ although the remainder of the calculation
involves a substantial amount of tedious algebra. We find
\begin{eqnarray}
\eta_3 &=& \eta^3_1 \left[ \mu \Theta \left( \frac{(3\mu-1)}{4(\mu-1)^2}
+ \frac{\mu(\mu-16)}{4(\mu-1)^2} - \frac{1}{4\mu} + \frac{6\mu}{(\mu-1)^2}
\right) \right. \nonumber \\
&+& \left. \frac{(2\mu-1)^2\Phi}{2(\mu-1)^2} + \frac{3\mu^2\Xi}{2(\mu-1)^3}
+ \frac{3\mu^2\Theta\Xi}{2(\mu-1)} + \frac{3\mu^2\Theta\Psi}{(\mu-1)}
+ \frac{3(2\mu-1)^2\Psi^2}{2(\mu-1)^2} \right. \nonumber \\
&+& \left. \Psi \left( \frac{\mu^3}{(\mu-1)^3} - \frac{\mu(2\mu-1)}{2(\mu-1)^3}
+ \frac{\mu^2(2\mu^2-5\mu+4)}{2(\mu-1)^3} - 2\mu - 3 + \frac{3}{2\mu}
\right. \right. \nonumber \\
&+& \left. \left. \frac{35}{2(\mu-1)} + \frac{18}{(\mu-1)^2}
+ \frac{9}{2(\mu-1)^3} \right) + \frac{1}{2\mu^2} - 4 - \frac{3}{\mu}
- \frac{7}{(\mu-1)} \right. \nonumber \\
&+& \left. \frac{13}{2(\mu-1)^2} + \frac{8}{(\mu-1)^3} + \frac{2}{(\mu-1)^4}
- \frac{\mu^2(2\mu-1)(\mu-3)}{(\mu-1)^3} \right]
\end{eqnarray}
which is the main result of this paper and represents the first $O(1/N^3)$
analysis in this model in arbitrary dimensions. (Previous $O(1/N^3)$
analysis was strictly two dimensional and examined the corrections to the
$\sigma$ field mass, \cite{34}.) We can evaluate (5.1) in three dimensions
to obtain
\begin{equation}
\eta_3 ~=~ \frac{256}{27\pi^6} \left[ \frac{47\pi^2}{12} + 9\pi^2\ln2
- \frac{189\zeta(3)}{2} - \frac{167}{9} \right]
\end{equation}
which has a similar structure to the analogous quantity in the $O(N)$ bosonic
$\sigma$ model in terms of the appearance of numbers such as $\ln 2$ and
$\zeta(3)$.

We conclude with several remarks. One important point of the analysis we have
described here is the adaptation of conformal techniques and the conformal
bootstrap programme to a model with fermions. Whilst the interaction is
relatively simple it ought now to be possible to perform an analogous
calculation in a gauge theory such as QED where the only essential
difference is the appearance of a $\gamma$-matrix at each vertex. This is
partly motivated by the fact that $\eta_2$ could be deduced via the self
consistency approach of \cite{10} in \cite{35} and it ought therefore to
be possible to construct analogous bootstrap equations for that model
to allow it to be probed as far as is analytically possible.

\vspace{1cm}
\noindent
{\bf Acknowledgement.} The author thanks Dr J.R. Honkonen for a useful
conversation.

\vspace{1cm}
\noindent
{\bf Note added.} Whilst in the final part of this work we received a
preprint, \cite{36}, where $\eta_3$ is also recorded and we note that both
results are in agreement.

\newpage
\noindent
{\Large {\bf Figure Captions.}}
\begin{description}
\item[Fig. 1.] Uniqueness rule for a fermionic vertex.
\item[Fig. 2.] Conformal triangle.
\item[Fig. 3.] Expansion of $3$-vertex.
\item[Fig. 4.] Regularized $3$-vertex graph.
\item[Fig. 5.] Intermediate graphs in the evaluation of $\Gamma_2$.
\item[Fig. 6.] $\Gamma_4$ after several integrations.
\end{description}
\end{document}